\documentclass[runningheads]{llncs}

\usepackage{graphicx}
\usepackage{hyperref}
\usepackage{amsmath, amssymb}

\usepackage[most]{tcolorbox}

\usepackage{graphicx}
\usepackage{caption}
\usepackage{subcaption}
\usepackage{float}

\begin{document}

\title{Oracle Counterpoint: Relationships between On-chain and Off-chain Market Data}
\titlerunning{Oracle Counterpoint}

\author{Zhimeng Yang\textsuperscript{1} \and Ariah Klages-Mundt\textsuperscript{2} \and Lewis Gudgeon\textsuperscript{3}}
\institute{Coinbase \and Cornell University \and Imperial College London}

\authorrunning{Yang et al.}

\maketitle

\begin{abstract}
We investigate the theoretical and empirical relationships between activity in on-chain markets and pricing in off-chain cryptocurrency markets (e.g., ETH/USD prices).
The motivation is to develop methods for proxying off-chain market data using data and computation that is in principle verifiable on-chain and could provide an alternative approach to blockchain price oracles.
We explore relationships in PoW mining, PoS validation, block space markets, network decentralization, usage and monetary velocity, and on-chain Automated Market Makers (AMMs).
We select key features from these markets, which we analyze through graphical models, mutual information, and ensemble machine learning models to explore the degree to which off-chain pricing information can be recovered entirely on-chain.
We find that a large amount of pricing information is contained in on-chain data, but that it is generally hard to recover precise prices except on short time scales of retraining the model.
We discuss how even noisy information recovered from on-chain data could help to detect anomalies in oracle-reported prices on-chain.
\end{abstract}


\section{Introduction}

Decentralized finance (DeFi) aims to transfer the role of trusted but risky intermediaries to more robust decentralized structures.
A remaining weak link is in reliance on off-chain information, such as prices of reference assets, which need to be imported on-chain through oracles.
The issue is that oracle-reported prices cannot be proven on-chain because the price process (usually in USD terms) is not observable there.

Various oracle security models exist, as described in \cite{werner2021sok}, though for the most part, they always involve some sort of trusted party or medianizing of several trusted parties. Even alternatives like referencing time weighted average prices (TWAPs) on decentralized exchanges (DEXs) still essentially involve a trusted party. In particular, to price an asset in USD terms, the standard approach is to use a DEX pair with a USD stablecoin, which shifts the trusted party to the stablecoin issuer as a sort of oracle.
While real trades can be observed on-chain, which is a strength of this method, stablecoin issuers can become insolvent, censor transactions, or otherwise cause the stablecoin to depeg. A resilient oracle system would ideally handle the corner cases of stablecoin pricing.

In this paper, we explore a new direction in oracle design wherein an \emph{estimate} of an off-chain price can in principle verifiable on-chain, which is intended to be used in addition to other oracle methods.
We investigate the theoretical and empirical relationships between activity in on-chain markets and the overall pricing and liquidity in off-chain cryptocurrency markets (e.g., \{BTC, ETH\}/USD price. 
The motivation is to develop methods for proxying off-chain market data using data within an on-chain environment.

\begin{figure}
	\centering
	\begin{subfigure}{0.75\textwidth}
			\centering
			\includegraphics[width=\textwidth]{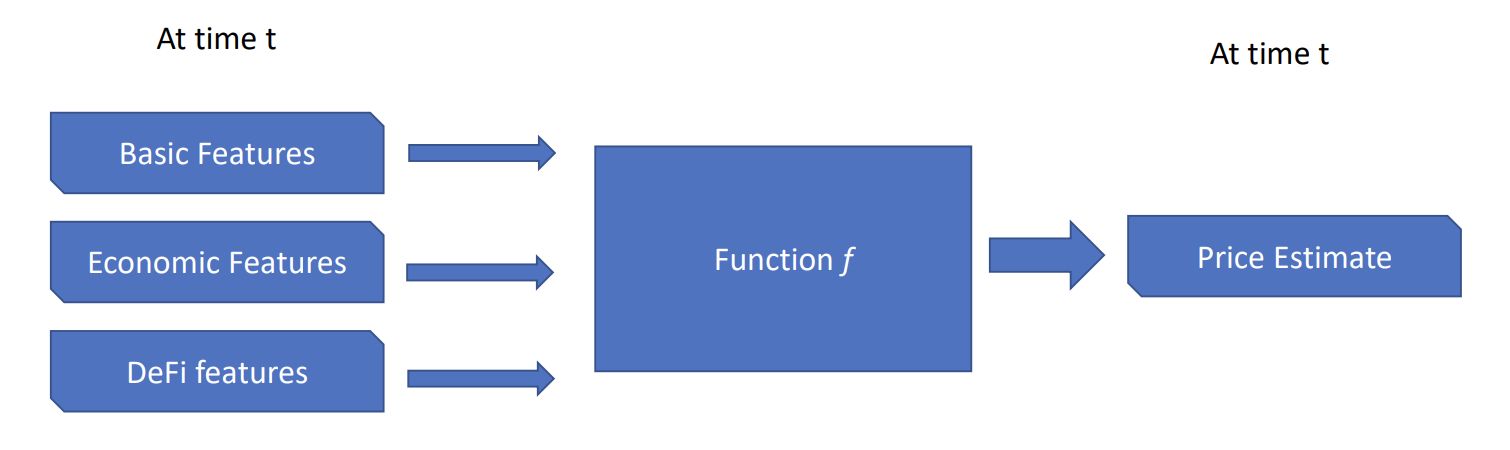}
			\caption{}
			\label{fig:process-chart}
		\end{subfigure}%
	\begin{subfigure}{0.25\textwidth}
			\centering
			\includegraphics[width=\textwidth]{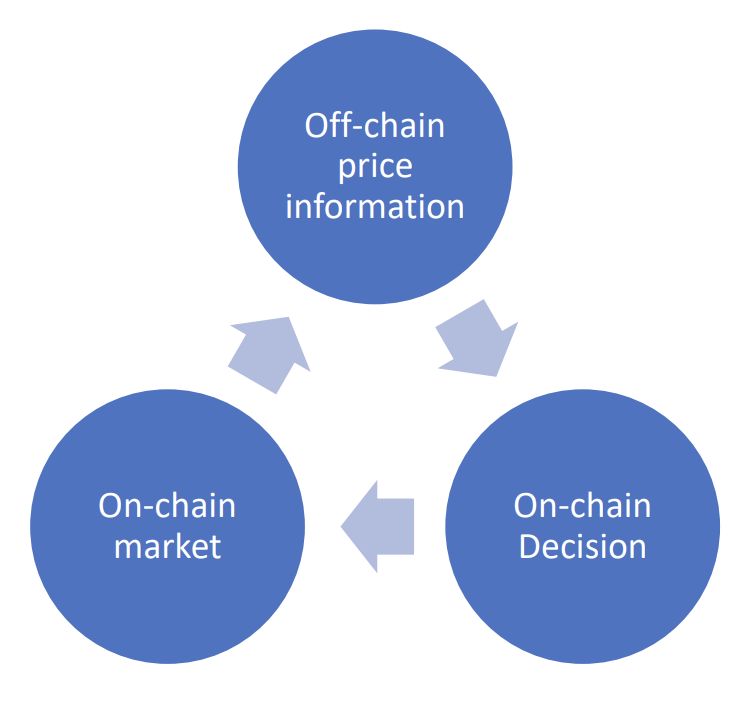}
			\caption{}
			\label{fig:hypothesis-causality}
		\end{subfigure}
	\caption{Proposed structure to estimate prices verifiably on-chain.}
	\label{fig:intro-charts}
\end{figure}

We formalize this as the task of finding a function $f$ that maps on-chain observable data (or features for the machine learning model) to close estimates of off-chain prices, as visualized in Figure~\ref{fig:process-chart}.
We consider a variety of features: \emph{basic features} that are straightforward properties about the network, \emph{economic features} that are computed from basic features and are suggested by fundamental economic models of the network, and features that arise from the usage of DeFi protocols.
Ideally, a good $f$ will also have two further properties: (i) it is difficult/costly to manipulate the output of $f$ through manipulating the inputs, and (ii) outputs of $f$ are provable on-chain.
The hypothesis predicating this structure is that off-chain price data (e.g., in USD terms) is incorporated into the behavior of agents in on-chain markets (e.g., mining, block space, and DeFi markets) and that
on-chain data thus provides some information that can be recovered about the original off-chain prices, as visualized in Figure~\ref{fig:hypothesis-causality}.

To understand the problem intuitively, compare with the usual financial price prediction problem, in which we would try to identify several drivers of future price and formulate a model to predict future prices with these drivers as features. The problem we consider is the reverse in some ways. In particular, we hypothesize that the price is a driving factor (probably one of many) behind the behavior of agents in on-chain markets, and we want to recover the current period price from the current state of on-chain market behaviors as features.

We explore this problem using a combination of economic theory about on-chain markets and data-driven analysis to explore the degree to which off-chain pricing information can be recovered from on-chain data. We find a meaningful price signal is recoverable as well as several strong empirical relationships with on-chain features. While it is not precise enough to use directly as an oracle, we discuss ways in which it could be used as a trustless sense check for oracle-reported prices. We finish by discussing several significant challenges that remain in developing and executing such a tool.
\section{Methods}

We explored relationships between off-chain ETH/USD and CELO/USD\footnote{ETH/USD is the focus as it is the main oracle input for DeFi protocols and is explored for PoW Ethereum data. CELO/USD is included as a first look at PoS data.} prices and feature variables from PoW mining for Ethereum and Bitcoin (alternatively PoS validation for Celo), block space markets, network decentralization (e.g., burden on running a full node), usage and monetary velocity, and DeFi liquidity pools and AMMs, including activity on Bitcoin, Ethereum, and Celo networks. We obtained raw block and transaction data from Google Cloud Bigquery, Uniswap v1 and v2 data from the Graph, and off-chain USD price data from the Coinbase Pro API. We then derived the following types of on-chain data features:
\begin{itemize}
	\item \emph{Basic network features} that can be straightforwardly derived from Ethereum block and transaction data, covering information related to Ethereum’s network utility, ether supply,  transaction cost and the network’s computational consumption (i.e. the gas market).
	\item \emph{Uniswap features} on participation in DEX pools involving ETH and stablecoins (DAI, USDC, USDT). For the most part, we intentionally do not focus on DEX prices, as those measures would equivalently treat the stablecoin issuer as a sort of trusted oracle. We instead mainly focus on a measure of liquidity moving in and out of DEX pools.
	\item \emph{Economic features} that are suggested by fundamental economic models of decentralized networks and can also be derived from on-chain data (as described in the next subsection).
\end{itemize}
Data was collected spanning from July 1 2016 to May 1 2022 and was aggregated to the hourly level. We include Bitcoin data along with Ethereum data in the dataset for the sake of exploring relationships as in principle it can also be verified on-chain to varying degrees and discuss the connections further later.

Some further details on data and features are provided in the appendix. Precise methods are available in the project github repo: \url{https://github.com/tamamatammy/oracle-counterpoint}.

\subsection{Fundamental Economic Features from On-chain Markets}

In addition to the above raw on-chain features, we also considered transformations of these features informed by fundamental economic models of on-chain markets, including PoW mining, PoS validation, block space markets, network decentralization costs of running full nodes, usage and monetary velocity, and on-chain liquidity pools (e.g., \cite{Huberman2019AnSystem,Prat2017AnMining,SusanAtheyIvaParashkevovVishnuSarukkai2016BitcoinUsage,Buterin2018BlockchainPricing,Fanti2019EconomicsINCOMPLETE,Easley2018FromHttps://ssrn.com/abstract=3055380}).
We analyzed the structure of these models to extract features that should economically be connected to price.

For example, \cite{Huberman2019AnSystem} models a block space market and finds that the ratio of average demand to capacity $\rho = \frac{\lambda}{\mu K}$ plays an important role in linking users' waiting costs to transaction fees pricing. Here $\lambda$ is the transaction volume, $K$ is the maximum number of transactions in a block, and $\mu$ is the block adding rate. A function emerges, called $F(\rho)$ that describes the relationship between fee pricing and congestion (i.e., amount of demand compared to supply for block space), which can be translated as
$$\text{tx fees in USD} = (\text{tx fees in ETH}) * \text{price}_{ETH} = F(\rho).$$
While $F(\rho)$ is nontrivial to work with, various pieces of the results in \cite{Huberman2019AnSystem} can be incorporated into useful features for the task of recovering $\text{price}_{ETH}$ (i.e., ETH/USD), including $\rho$, $\rho^2$, and the empirical finding that $\rho=0.8$ represents a phase transition in fee market pricing.

We also used the model in \cite{SusanAtheyIvaParashkevovVishnuSarukkai2016BitcoinUsage}, which modeled cryptocurrency price based on market fundamentals. A key feature in their model was currency velocity, which is defined as the ratio of transaction volume to money supply:
$$ \text{Velocity} = \frac{\text{Transaction Volume (USD)}}{\text{Exchange Rate} * \text{Supply of Currency}}.  $$
In their model, they show a unique equilibrium cryptocurrency exchange rate based on supply and demand with steady state expected exchange rate equal to the ratio of expected transaction volume and cryptocurrency supply.
Based on this model, we also incorporated the ratio of transaction volume (in cryptocurrency units since USD value is not given) and cryptocurrency supply, measured over one hour time steps, as an additional feature variable in our analysis.

We formulate other factors related to mining payoff, computational burden, and congestion as reviewed in the appendix.

\subsection{Data-driven Feature Analysis}\label{sec:feature-analysis}

We analyse empirical relationships between features using graphical models and mutual information to study which features are most related to USD prices.

We use probabilistic models of Markov random fields, generated through sparse inverse covariance estimation with graphical lasso regularisation over normalized data, to express the partial/pair-wise relations between the time series of on-chain feature variables and off-chain prices.
For example, if the true underlying structure is Gaussian, then entries of the inverse covariance matrix are zero if and only if variables are conditionally independent (if the underlying structure is not Gaussian, then in general the inverse covariance matrix gives partial correlations).\footnote{In comparison, covariance represents relations between all variables as opposed to partial/pairwise relations.}
Sparse inverse covariance estimation is a method for learning the inverse covariance structure from limited data by applying a form of L1 regularization (see \cite{friedman2008sparse}).

The output of this technique helps to uncover strong empirical dependencies within the data, suggesting features that are strongly related to price and others that replicate similar information as others. We find that the method is often sensitive to the precise dataset used, which we adjust for by smoothing over the outputs of many $k$-fold subsets.

We also consider mutual information between features in the dataset, which describes the amount of information measured (in information entropy terms), measured in reduction of uncertainty, obtained about price by observing the on-chain features. In information theory, entropy measures how surprising the typical outcome of a variable is, and hence the `information value'. This is helpful both in identifying strong relationships and evaluating different smoothing factors considering noisy on-chain signals. In this analysis, we consider smoothed versions of the feature set based on exponential moving averages with memory parameters $\alpha$, i.e., for feature value $b_t$ at time $t$, the smoothed measure is
$$ \tilde b_t = (1-\alpha) b_t + \alpha \tilde b_{t-1}.$$

\subsection{Modeling Off-chain Prices}

We apply supervised machine learning methods to explore the degree to which off-chain pricing information can be recovered from information that is entirely on-chain. We apply a few select simple and ensemble supervised machine learning methods on a rolling basis: basic regression, single decision tree, random forest, and gradient boost. 
A decision tree is a non-parametric method that breaks down the regression into subcases of the feature variables following a tree structure. A tree ensemble method constructs many tree models and uses the average of the trees as the output.
The motivation for using tree-based ensemble methods is the non-parametric nature of the dataset and success of similar methods in analyzing other market microstructure settings \cite{easley2021microstructure}.
In our results, we focus on random forest, which trains many trees to different samples of the empirical distribution of training points, as this method tends to be resilient to overfitting.

\begin{figure}
	\centering
	\includegraphics[width=\textwidth]{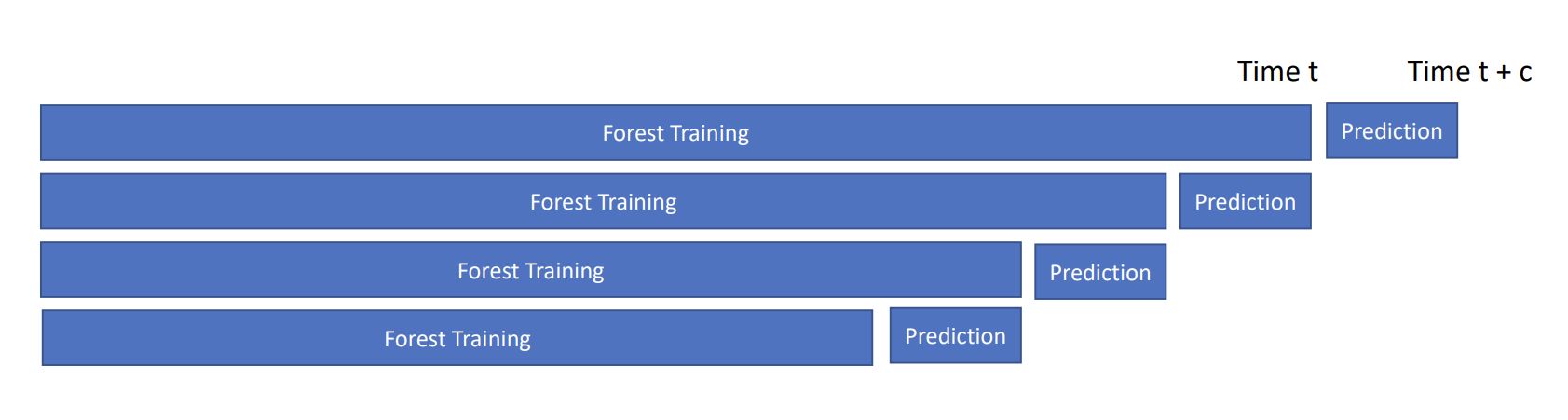}
	\caption{Rolling training-testing data split}
	\label{fig:model-training}
\end{figure}

We run these models on the data set and evaluate performance using out-of-sample testing data on a rolling basis. The rolling training-testing data split, as depicted in Figure~\ref{fig:model-training}, is applied to boost model performance. For a given set of time series data with time duration of time t + time c = time t+c, where time series before time t were used for model training and time series between time t and time t + c were used for model testing. The benefit of this split is to test how good the model is in proxying ETH USD price for a fix period in the future, with all the information available in the past.

\section{Results}

We focus on Ethereum data analysis under PoW in this section. Analysis of Celo data is included in the appendix as a first look at a PoS system. There is not yet enough historical data to analyze Ethereum PoS but would be a next step.

\subsection{Feature Analysis}

We find that a large amount of off-chain pricing information is contained in on-chain data and that the various features are connected in some strong but complicated ways.

\begin{figure}
	\centering
	\includegraphics[width=0.8\textwidth]{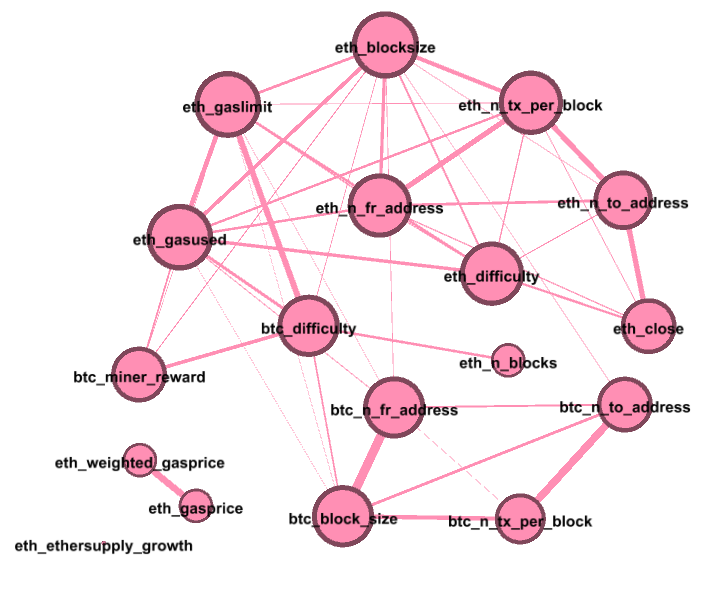}
	\caption{Graphical network visualization. Variables defined in Table~\ref{table:feature-defs}.}
	\label{fig:graphical-model}
\end{figure}

Figure~\ref{fig:graphical-model} and Appendix Figure~\ref{fig:partial-corr} show the results of sparse inverse covariance modeling for a selection of the feature set. 
The graphical structure depicted is the consistent structure over time as smoothed over the outputs of many $k$-fold subsets. The partial correlation matrix shows the graphical structure in matrix form.
The features that are most directly connected with ETH/USD price, as measured by partial correlations in the graphical model, include number of active to and from addresses sending transactions, block difficulty, and number of transactions per block.
Several of these variables appear to contain information relevant to price as well, as measured by mutual information in the next analysis. The graphical model suggests that these are indirect relationships, though it is worth noting that the process is unlikely Gaussian, and so the partial correlations do not necessarily translate to conditional dependencies.

\begin{figure}
	\centering
	\includegraphics[width=\textwidth]{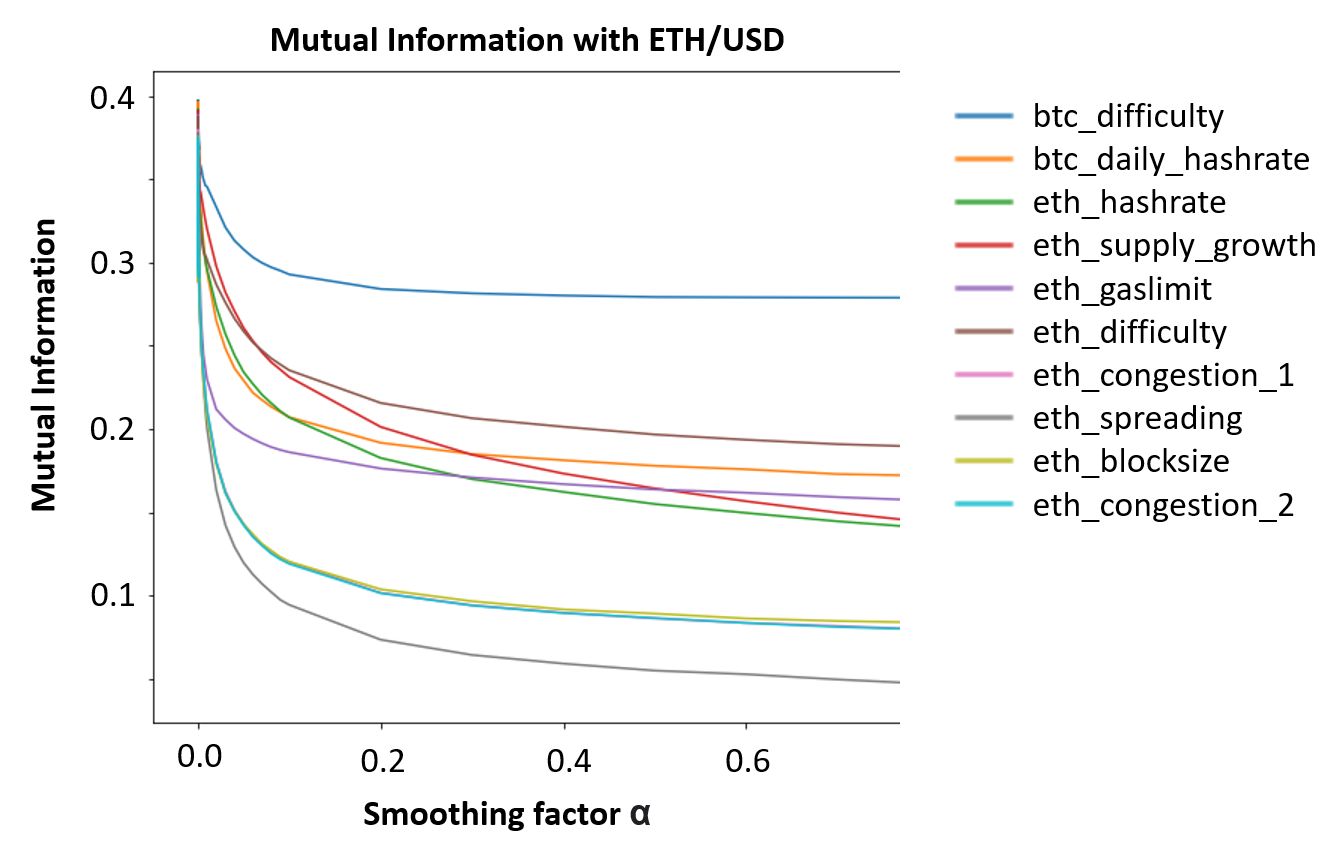}
	\caption{Mutual information of price data and feature variables, with memory parameter (or smoothing factor) $\alpha$ applied to the feature variables (see \ref{sec:feature-analysis}). Variables defined in Table~\ref{table:feature-defs}.}
	\label{fig:mutual-info}
\end{figure}

Figure~\ref{fig:mutual-info} shows the mutual information between ETH/USD prices and other features, meaning the amount of information (reduction of uncertainty) obtained about price by observing each other variable individually. We find that across the top 10 features, a large amount of information about off-chain price is contained in on-chain data. We also find that the mutual information decreases with $\alpha$, the exponential moving average memory factor for smoothing, indicating that the smoothed data is generally less informative than the most up-to-date data.

We also analyze the full feature set, including the transformed economic factors and Uniswap pool liquidity factors. Perhaps unsurprisingly, since the transformed features contain the same underlying information, they do not exhibit stronger relationships than the raw features. More surprising is that the Uniswap pool factors also did not present strong relationships with price. We then arrived at the above version of the analysis excluding Uniswap factors enabling us to use the entire data history (as Uniswap was launched later than the start of the dataset).

\subsection{Recovering Off-chain Prices from On-chain Data}

Random forest and gradient boost both outperformed the other two simpler ML algorithms. We selected Random Forest as the candidate model in the end as it is in principle simpler to be implemented on-chain compared to the gradient boost model (theoretically, a random forest model could be implemented as one big mapping table in a smart contract).

We tested the model performance over different lengths of period - the length of time duration between time t and time t+c. As would be expected with nonstationary time series, we observed that the longer the time duration that a single trained model is used for price estimation, the less accurate is price estimation. The degree to which time between retrainings affects accuracy is informative, however.

\begin{figure}
	\centering
	\includegraphics[width=\textwidth]{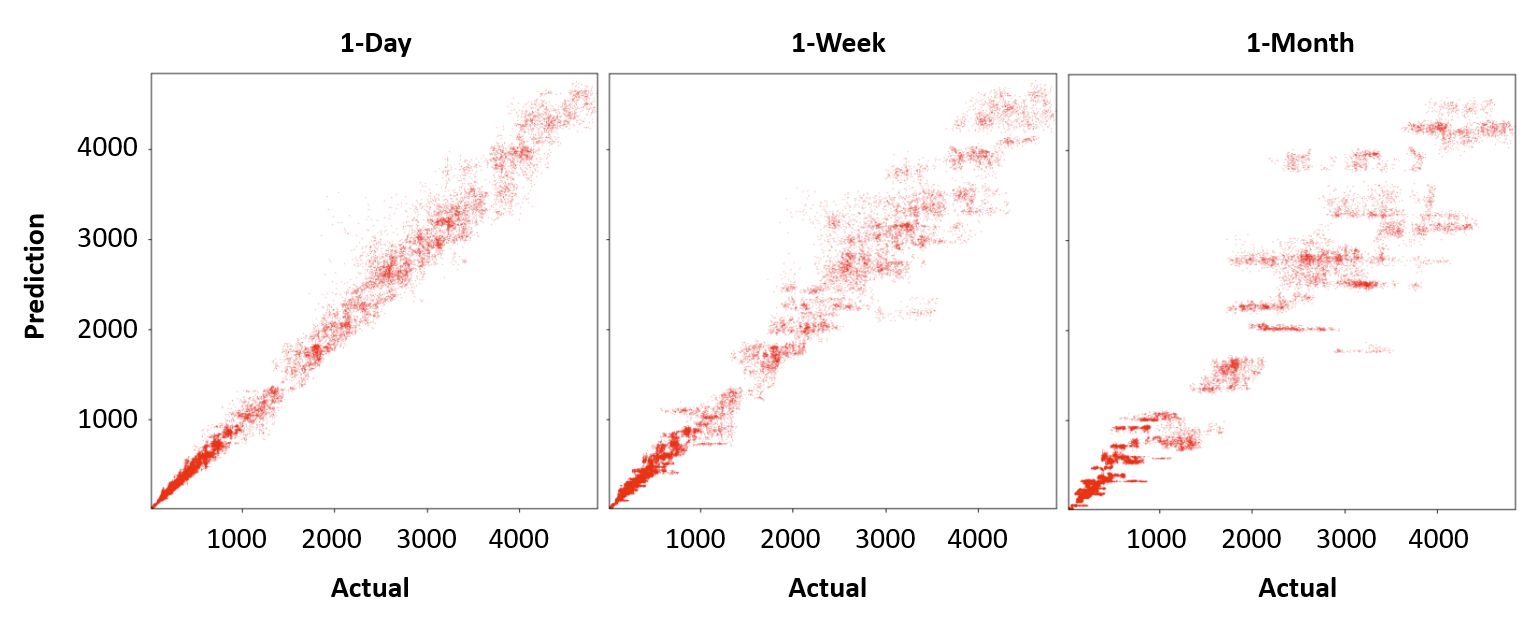}
	\caption{Recovered price vs actual for random forest with given retraining periods.}
	\label{fig:pred-retrainings}
\end{figure}

Figure~\ref{fig:pred-retrainings} shows the random forest model performances, Estimated vs Actual ETH/USD price, for 1-day ahead, 1-week ahead and 1-month ahead of retrainings. While none of the models provide high accuracy of recovering ETH prices, they do demonstrate that a good signal of the general price level can be recovered, particularly in the 1-day and somewhat in the 1-week retraining cases.

The deviation between estimated price and actual price is bigger for higher ETH prices. This is a combination of both having less data in the dataset for these prices and the fact that the same relative error scales with the absolute price, and so deviations measured absolutely are expected to be greater.

We run the models on the full feature set, including transformed economic factors and Uniswap pool factors. The economic factors provide little new information vs the raw features, perhaps a consequence of the flexibility of the tree models. Uniswap pool factors similarly do not improve accuracy. The final analysis excludes Uniswap factors enabling the entire data history to be used.

\subsection{Performance of Price Recovery}\label{sec:performance}

To measure performance of the price recovery models, we compare against a simple martingale benchmark. This benchmark supposes that the last observed price in the last retraining period is the best estimate of the next price in expectation, barring any new information, which would follow in an efficient market. By comparing against this benchmark, we evaluate how well the on-chain feature variables, the only source of new information to the model, recover price vs the best guess without this information.

We evaluate the squared error between a prediction (either the model or benchmark price) at time $t$ and the actual price as
$$SE = (\text{predicted}/\text{actual} -1)^2.$$
We then consider the mean squared error (MSE) over different times $t$. The square root of the MSE (RMSE) then gives a measure of error that can be interpreted as a percentage of the price level.

We compare model errors with benchmark errors using these measures. We first consider the difference in squared errors between the model and the benchmark as
$$DSE = (\text{benchmark}/\text{actual}-1)^2 - (\text{model}/\text{actual}-1)^2.$$
This quantity will be positive when the model performs better than the benchmark and negative otherwise.
Table~\ref{table:DSEs} summarizes how frequently the models have lower squared error than the benchmarks and by how much the squared error is reduced (as a percentage of price level) when this happens. Note that most of the time, the models perform worse than the benchmarks over the dataset. However, they may be able to provide useful information in addition to the benchmarks in some settings.

\begin{table}
	\centering
	\begin{tabular}{c | c  c  c}
		\textbf{Model retraining periods:} & \textbf{1-day} & \textbf{7-day} & \textbf{30-day} \\
		\hline
		How often model beats benchmark & 12.4\% & 26.9\% & 32.4\% \\
            Gain over benchmark when model is better & 0.65\% & 3.56\% & 7.10\% \\
		\hline
	\end{tabular}
	\caption{Summary of DSEs between models and benchmarks for different retraining periods evaluated on the whole dataset (2016-2022). Row 1 is the frequency that $DSE > 0$. Row 2 is the root mean DSE at the times that $DSE > 0$.}
	\label{table:DSEs}
\end{table}

In line with Table~\ref{table:DSEs}, the MSE of the models is greater than the MSE of the benchmarks when taken over the entire dataset. A limitation with this measure is that, for data points early in the dataset, there is relatively less training data for the rolling models. We might expect the models to do better toward the end of the dataset where there is more training data to work with.

In Table~\ref{table:RMSEs}, we calculate RMSEs restricted to the last year of the dataset, when the models can theoretically be best. In addition to RMSE over this time period, we also compute RMSE during the top 10\% most volatile days, as measured by rolling 24 hour volatility calculated on hourly returns.

Table~\ref{table:RMSEs} shows some limited situations where a model makes improvements over the benchmark as measured in less RMSE compared to the benchmark. This happens for the 30-day retraining model, which also tends to perform better for the most volatile days. In general, the model error is usually larger than the benchmark error, however, and the outperformance of the 30-day model is somewhat sensitive to the restriction to the final year, which is further explored in Appendix~\ref{appendix:performance}.

\begin{table}
	\centering
	\begin{tabular}{c | c  c  c}
		\textbf{Model retraining periods:} & \textbf{1-day} & \textbf{7-day} & \textbf{30-day} \\
		\hline
		Model RMSE & 7.82\% & 18.83\% & 18.98\% \\
            Benchmark RMSE & 3.77\% & 9.39\% & 19.80\% \\
            \hline
            Model (top 10\% vol) RMSE & 15.41\% & 23.15\% & 29.84\% \\
            Benchmark (top 10\% vol) RMSE & 7.5\% & 12.13\% & 36.61\% \\
		\hline
	\end{tabular}
	\caption{RMSEs of the models compared to benchmarks over the last year of the dataset (May 2021 - May 2022).}
	\label{table:RMSEs}
\end{table}

While the current models hint that on-chain information could be useful in reducing error, in practice they are not precise enough. In particular, an application could get most of the utility of the current models by checking that oracle prices don't change too quickly (i.e., implement a check of current oracle prices against the benchmark estimate of a previously observed price). Note that the last observed price in the last retraining period is not an input variable to the price recovery model other than as part of training. A next step in improving model performance could be to incorporate this last training price or to train the model to predict how current price deviates from the last training price. More refined methods could find ways of extracting better information from the on-chain data.

\section{Discussion}

We find that a general, but noisy, signal of off-chain prices can be extracted from the on-chain feature set, although it remains difficult to extract precise prices from the noise.
It is possible to improve the accuracy of the model by including features of DEX pricing of ETH/stablecoin pairs, as would be expected from \cite{angeris2020improved}. However, this would implicitly rely on the assumption that 1 stablecoin = 1 USD, which would face the significant further issues of detecting stablecoin depeg events (such as happened in USDC in March 2023) given that data is sparse for such events. Instead, the aim of this work is to provide information that can be used on top of existing oracle mechanisms, including DEX pricing, to relax trust requirements in those methods.

While this approach could likely not be used as a direct price oracle, the information from the recovered price signal could still in principle be useful as a sense check to inform when other oracle-reported prices may be suspect.
This function would be potentially very useful in application as the most profitable oracle manipulations to date have been large manipulations that may be caught by such methods.
An existing oracle system of this style in \cite{klages2022cpf} has been developed by cross-referencing information from DEX price sources. This approach has limitations, however, in sense checking the connection to the desired quote asset USD since it cannot be represented by DEX prices alone.
Incorporating measures of the price signal that we uncover on top of the existing structure could help to mitigate this limitation.
A main open question here is whether measures of price signal reported on-chain can be improved enough to make this feasible in application as a means of anomaly detection for oracle-reported prices.
Such a method could also serve to better align the incentives of an oracle provider to report correct prices with the knowledge that their quality of their feed is being graded against the signal in on-chain information. Models such as \cite{huo2021decentralized,klages2020stablecoins} could model this analytically, interchanging the oracle provider with the governors in those models.

Several challenges remain for implementing and running such a mechanism in practice.
One is accessing all the data within the EVM. Some of the data is in principle possible to access but may be too computationally intense under current systems. For instance, proving information about transactions or bridging BTC data might require running light clients on-chain. 
For example, see \cite{karantias2020smart,axiom} for possible methods of referencing normally difficult-to-obtain features of a chain from a smart contract.
For BTC data, this can mostly be ignored as it was not critical for the predictive models, but there was a lot of information in Ethereum transaction statistics.
It is worth noting that some features such as gas prices are easier to access now with EIP 1559.
Another challenge is in evaluating how manipulable the features are should a bad actor want to affect the price estimation. In principle, resilient measures seem possible considering that on-chain markets can be costly to manipulate, though may also be computationally burdensome to produce.

An implementation would also have to handle the rolling nature of retrainings required to accurately recover price data. The implementation would need a trust minimized way to update a smart contract implementation with new trainings.
In principle this is also possible, such as by implementing the training program in fixed point to run deterministically and implementing a way to prove the correctness of a training on-chain. However, this would be daunting from the technical side as well as likely costly to run in most environments. The burden could possibly be eased by running it `optimistically' by incorporating a challenge period and fraud proofs, though it's unclear if this would be enough of an improvement.
Another viable way is for a trusted trainer to regularly update calibrations on-chain subject to on-chain spot checks and not full proofs.

\subsection*{Acknowledgements.}
This paper is based on work supported by a Bloomberg Fellowship, EPSRC Standard Research Studentship (DTP) (EP/R513052/1) and a Celo grant.

\bibliographystyle{splncs04}

\appendix

\section{More Details on Dataset Features}

\begin{figure}[H]
	\centering
	\includegraphics[width=0.85\textwidth]{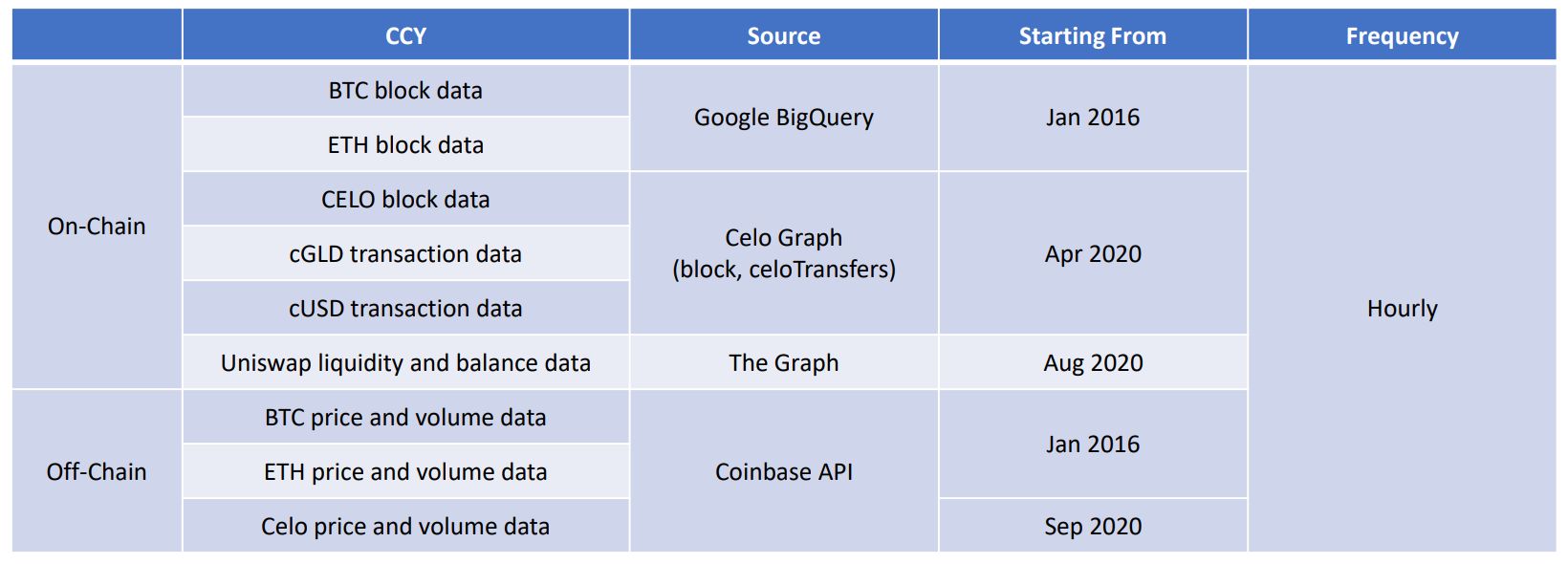}
	\caption{Overview of dataset.}
	\label{fig:dataset-overview}
\end{figure}

Figure~\ref{fig:dataset-overview} and Table~\ref{table:features} provide more information on the feature set used.

Table~\ref{table:features} describes the full feature variable set used at a high level, including basic network features, DeFi features from Uniswap, and features informed from economic models.

\begin{table}
	\centering
	\begin{tabular}{c | l}
		\textbf{Feature type} & \textbf{Feature (high level description)} \\
		\hline
		Network & Number of blocks\\
		 & Number of transactions \\
		 & \% change in accumulated ETH supply \\
		 & Avg gas limit \\
		 & Avg gas used \\
		 & Avg gas price \\
		 & Hash rate \\
		Uniswap & Liquidity in ETH/stablecoin pools \\
		& Trade volume in ETH \\
		Economic & Mining pay-off factors \\
		& Computational burden measures \\
		& Congestion factors \\
		& Social cost factors \\
		& Spreading factor \\
		\hline
	\end{tabular}
	\caption{Data features (high level).}
	\label{table:features}
\end{table}

Table~\ref{table:feature-defs} defines the variables used in the main text figures.

\begin{table}
\centering
\begin{tabular}{c | l}
    \textbf{Feature} & \textbf{Definition} \\
    \hline
    \texttt{eth\_gaslimit} & hourly average gas limit on Ethereum \\
    \texttt{eth\_gasused} & hourly average gas used per block \\
    \texttt{eth\_blocksize} & hourly average Ethereum blocksize (gas) \\
    \texttt{eth\_n\_from\_address} & hourly average sender addresses per block \\
    \texttt{eth\_n\_to\_address} & hourly average receiver addresses per block \\
    \texttt{eth\_supply\_growth} & hourly change in ETH supply \\
    \texttt{eth\_hashrate} & hourly average hashrate on Ethereum \\
    \texttt{eth\_gasprice} & hourly average tx gas price \\
    \texttt{eth\_weighted\_gasprice} & hourly avg tx gas price, weighted by gas used in tx / total gas used \\
    \texttt{eth\_congestion\_1} & hourly gas used / hourly gas limit \\
    \texttt{eth\_congestion\_2} &  square of \texttt{eth\_congestion\_1} \\
    \texttt{eth\_spreading} & hourly \# receiving addresses / hourly \# sender addresses \\
    \texttt{eth\_close} & hourly ETH/USD closing price \\
    \texttt{btc\_difficulty} & hourly average difficulty on Bitcoin \\
    \texttt{btc\_minter\_reward} & hourly average miner reward on Bitcoin \\
    \texttt{btc\_n\_from\_address} & hourly average sender addresses per block \\
    \texttt{btc\_block\_size} & hourly average Bitcoin block size (bytes) \\
    \texttt{btc\_n\_tx\_per\_block} & hourly average \# txs per block \\
    \texttt{btc\_to\_address} & hourly average receiver addresses per block \\
    \texttt{btc\_daily\_hashrate} & hourly Bitcoin difficulty / hourly average block time \\
    \end{tabular}
    \caption{Definitions of variables used in figures.}
    \label{table:feature-defs}
\end{table}

Online documentation in the project github repo provide further details of the underlying economic models and calculation of the economic factors (as well as calculation of other factors from the raw data): \url{https://github.com/tamamatammy/oracle-counterpoint}.

\subsection{Economic features}
A brief overview of the features informed by fundamental economic models is as follows along with citations for the relevant models that influenced the choice of these features.

\begin{itemize}
	\item Mining payoff factor 1: $(R (\text{blockReward} + \text{blockFees}))^{-1}$ \cite{Kroll2013TheAdversaries,Prat2017AnMining}
	\begin{itemize}
		\item R = block rate (/s), eth\_n\_blocks = \# blocks in the last hour
	\end{itemize}
	\item Previous high hash rate / current hash rate
	\item previous high $(R (\text{blockReward} + \text{blockFees}))^{-1} / \text{current}$
	\item Excess block space (block limit - gas used)
	\item Social value: D(W) is the social value of the level of decentralization = D(W) = - log(W) $\implies$ D(W) = - log(gas\_used) for ethereum, = - log(bytes); gas used as the measure of the weight of a block (W) \cite{Buterin2018BlockchainPricing}
	\item Social cost: Marginal cost = 1/gas\_used or 1/bytes \cite{Buterin2018BlockchainPricing}
	\item Computational burden on nodes: use block\_size as bandwidth $\implies$ $\text{block\_size} * \log^2(\text{block\_size})$ \cite{Buterin2018BlockchainPricing}
	\item Congestion factors: rho = gas used/gas limit, and $\text{rho}^2$; (in economic model, rho is defined as average number of transaction per block / number of transactions per block) \cite{Huberman2019AnSystem}
	\item Congestion factor: Indicator\_\{$\text{rho} > x$\}, heuristic use x = 0.8 \cite{Huberman2019AnSystem}
	\item Congestion pricing term 1: F(rho) / tx\_fees\_eth, where F describes relationship between USD tx fees and congestion \cite{Huberman2019AnSystem}
	\begin{itemize}
		\item Heuristic: use F = congestion factor 1 or 2 above
	\end{itemize}
	\item Congestion pricing term 2: max number of transactions in a block / fees in block \cite{Nicolas2014TheFees}
	\item Congestion pricing term 3: max number of transactions squared in a block / fees in block \cite{Nicolas2014TheFees}
	\item Spreading factor: number of unique output addresses / number of unique input addresses \cite{SusanAtheyIvaParashkevovVishnuSarukkai2016BitcoinUsage}
\end{itemize}

\section{Further Information on Ethereum Analysis}

Sparse inverse covariance estimation was performed with the implementation in SciPy using an alpha parameter of 4, convergence tolerance of 5e-4, 5 folds for cross valiation, 4 grid refinements, and 1000 max iterations.

\begin{figure}[H]
	\centering
	\includegraphics[width=\textwidth]{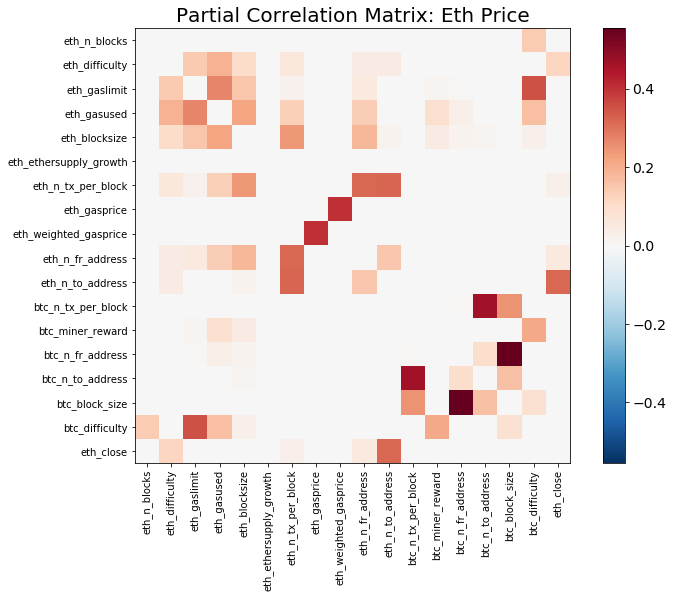}
	\caption{Partial correlation matrix from sparse inverse covariance estimation.}
	\label{fig:partial-corr}
\end{figure}

\subsection{Performance of Price Recovery}\label{appendix:performance}

\begin{figure}[H]
	\centering
	\includegraphics[width=\textwidth]{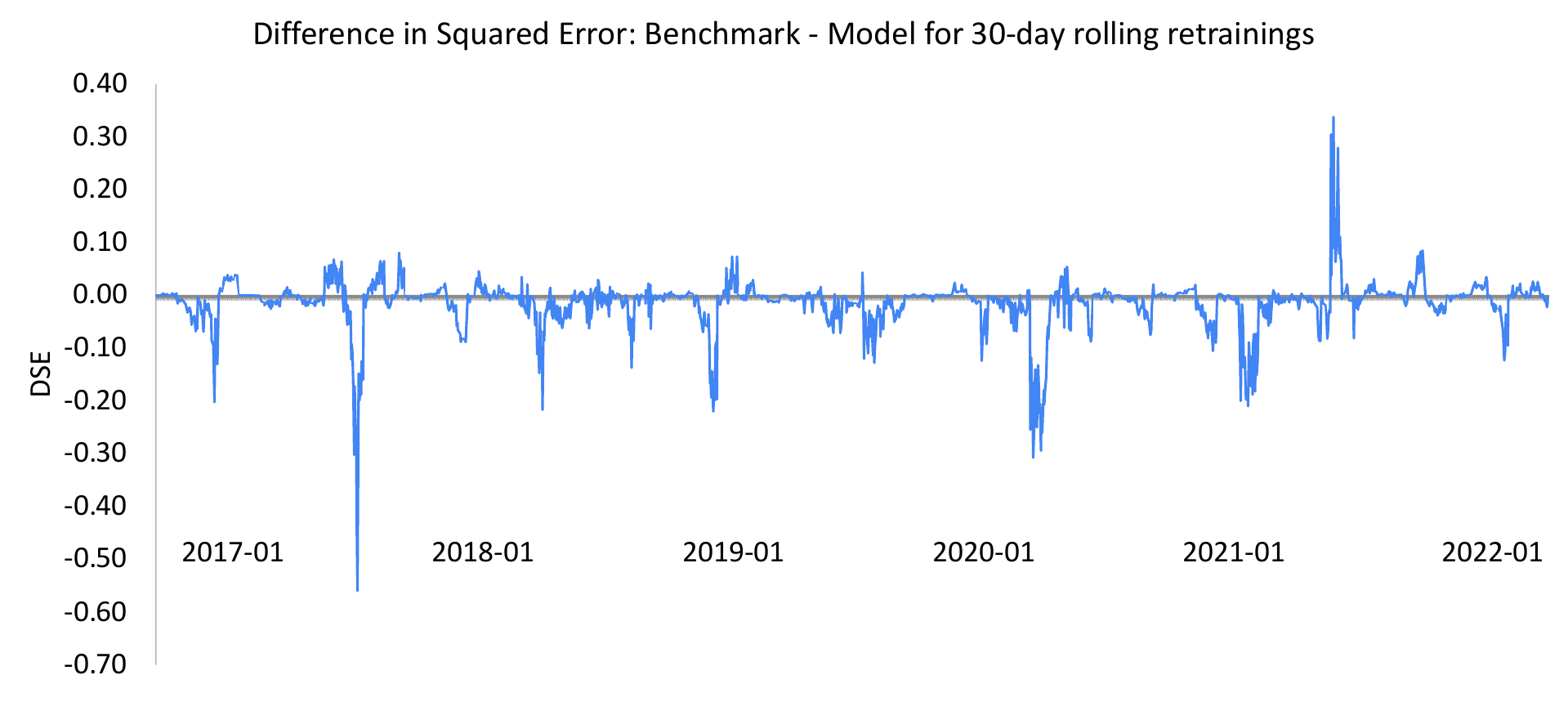}
	\caption{}
	\label{fig:DSEs}
\end{figure}

We continue the analysis from Section~\ref{sec:performance} with a few additional charts that give a wider view on performance for the 30-day rolling retraining model. Figure~\ref{fig:DSEs} shows the difference in squared error of the benchmark minus the model over time. When $DSE>0$, the model is performing better than the martingale benchmark of the last observed price in the rolling retraining. Of note is that while the performance improves in the final year (May 2021-May 2022), one small period represents most of the performance size.

\begin{figure}[H]
	\centering
	\includegraphics[width=\textwidth]{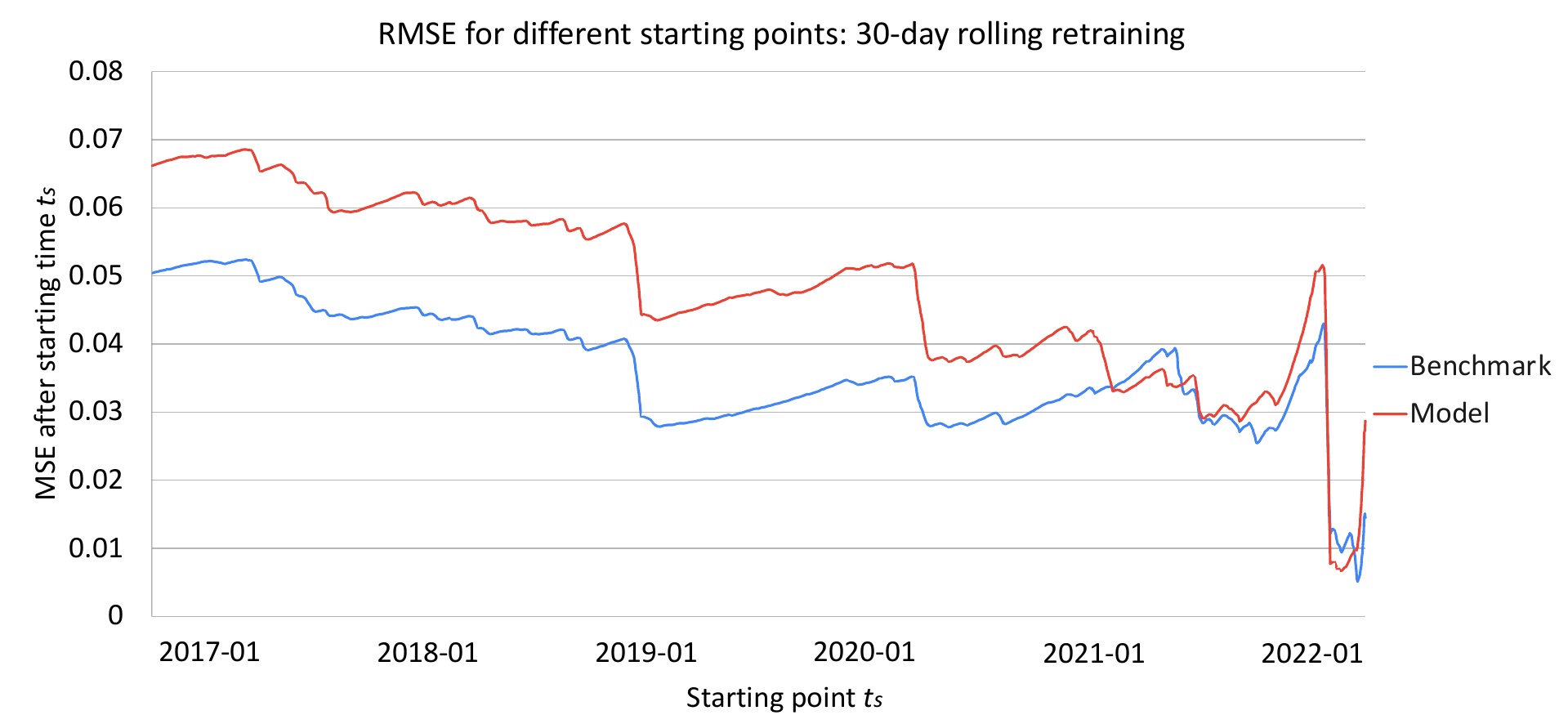}
	\caption{}
	\label{fig:RMSEs}
\end{figure}

Figure~\ref{fig:RMSEs} shows the RMSE evaluated from each different starting point on the x-axis to the end of the dataset (May 2022). As we move to later starting points on the x-axis, it is worth noting that there is more training data incorporated into the model before the test set for RMSE. Toward the end of the dataset, the model becomes more competitive with the benchmark, surpassing it when measured over the final year of data.

\begin{figure}[H]
	\centering
	\includegraphics[width=\textwidth]{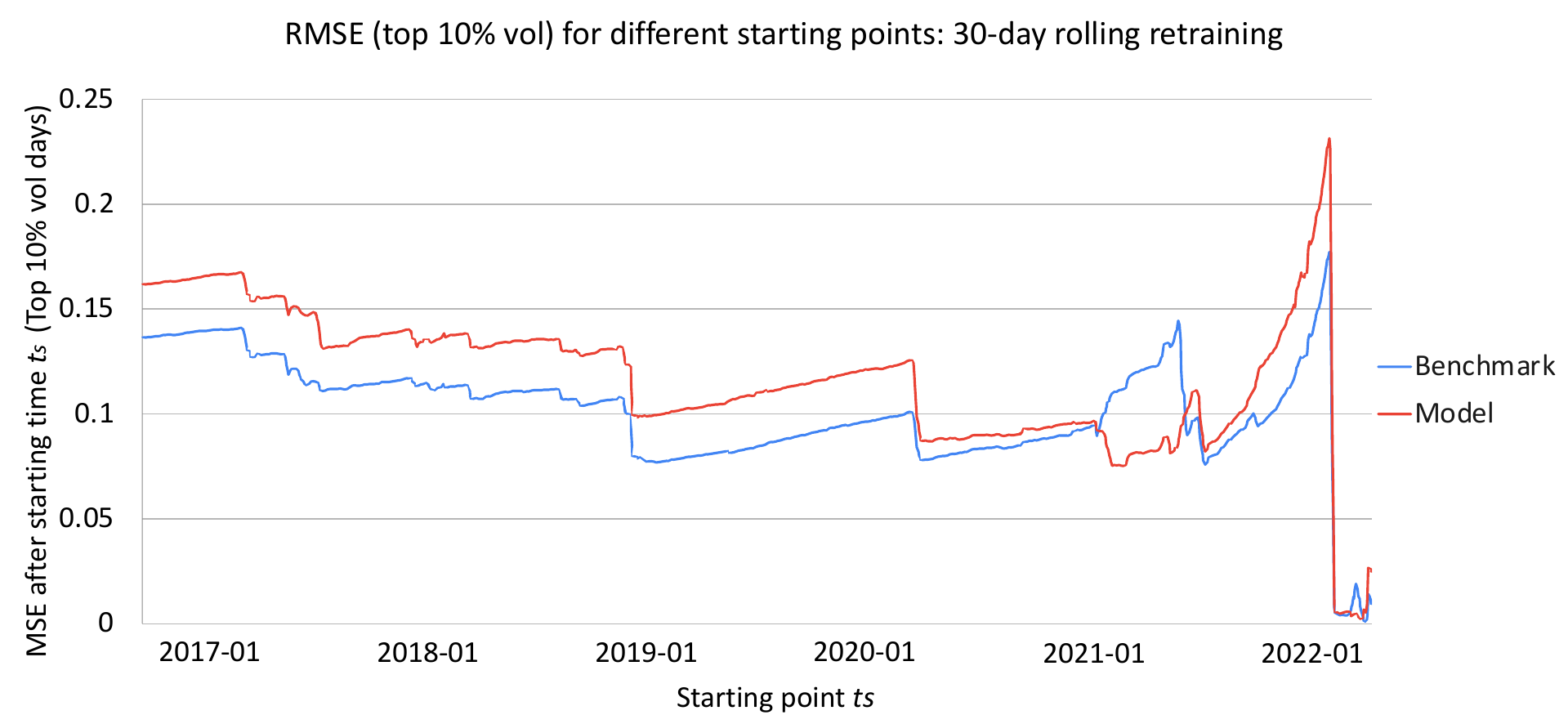}
	\caption{}
	\label{fig:RMSEs-vol}
\end{figure}

Figure~\ref{fig:RMSEs-vol} shows a similar plot of RMSE evaluated from different starting points on the x-axis, but with the calculation restricted to the top 10\% times of volatility. Here volatility is calculated as 24 hour rolling volatility of hourly returns. The model is overall more competitive with the benchmark for top 10\% volatility times compared to all times, and surpassing it by a sizable amount measured over the final year of data. Note that as suggested in Figure~\ref{fig:DSEs}, the outperformance of the model in the final year largely rests on a short period of high outperformance.

\section{Analysis of Celo PoS Data}

In addition to Ethereum data, we also analyse data on the Celo PoS network. This analysis involves some further features involving PoS systems as well as Celo's dual token model.
This additionally serves as a first look at the analysis of a PoS system with historical data spanning longer than a year. In comparison, a similar analysis of Ethereum's new PoS system does not yet have enough history at the current time to perform a good analysis.

\begin{figure}[H]
	\centering
	\includegraphics[width=\textwidth]{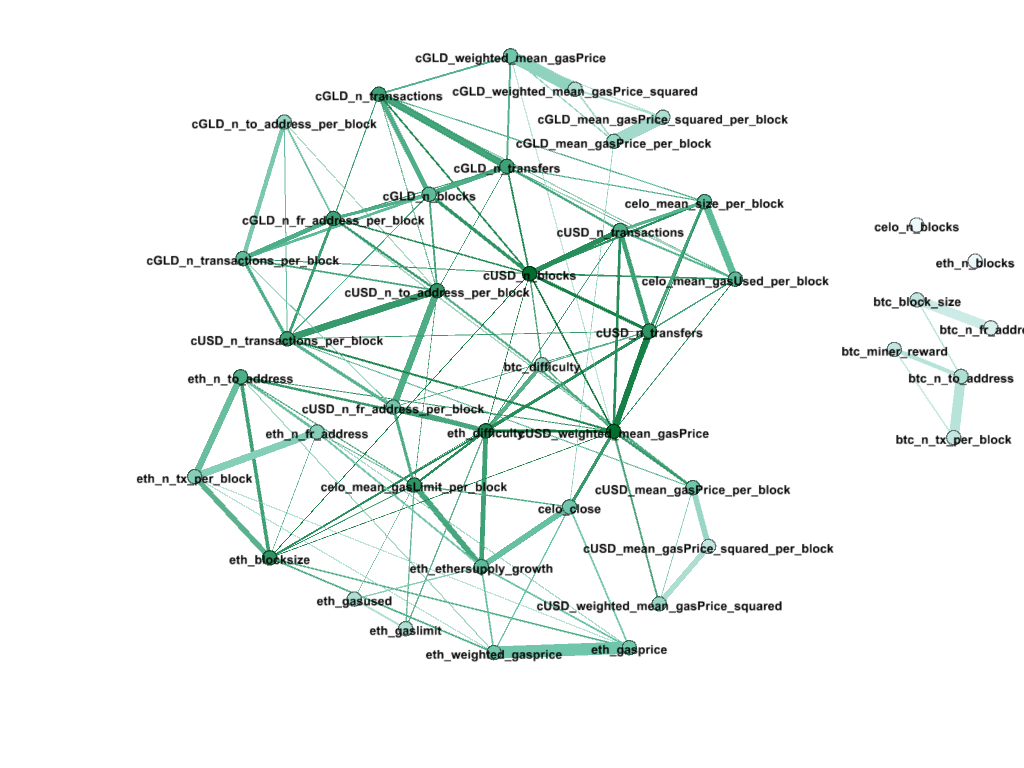}
	\caption{Graphical network visualization from sparse inverse covariance estimation.}
	\label{fig:partial-corr-graph-celo}
\end{figure}

\begin{figure}[H]
	\centering
	\includegraphics[width=\textwidth]{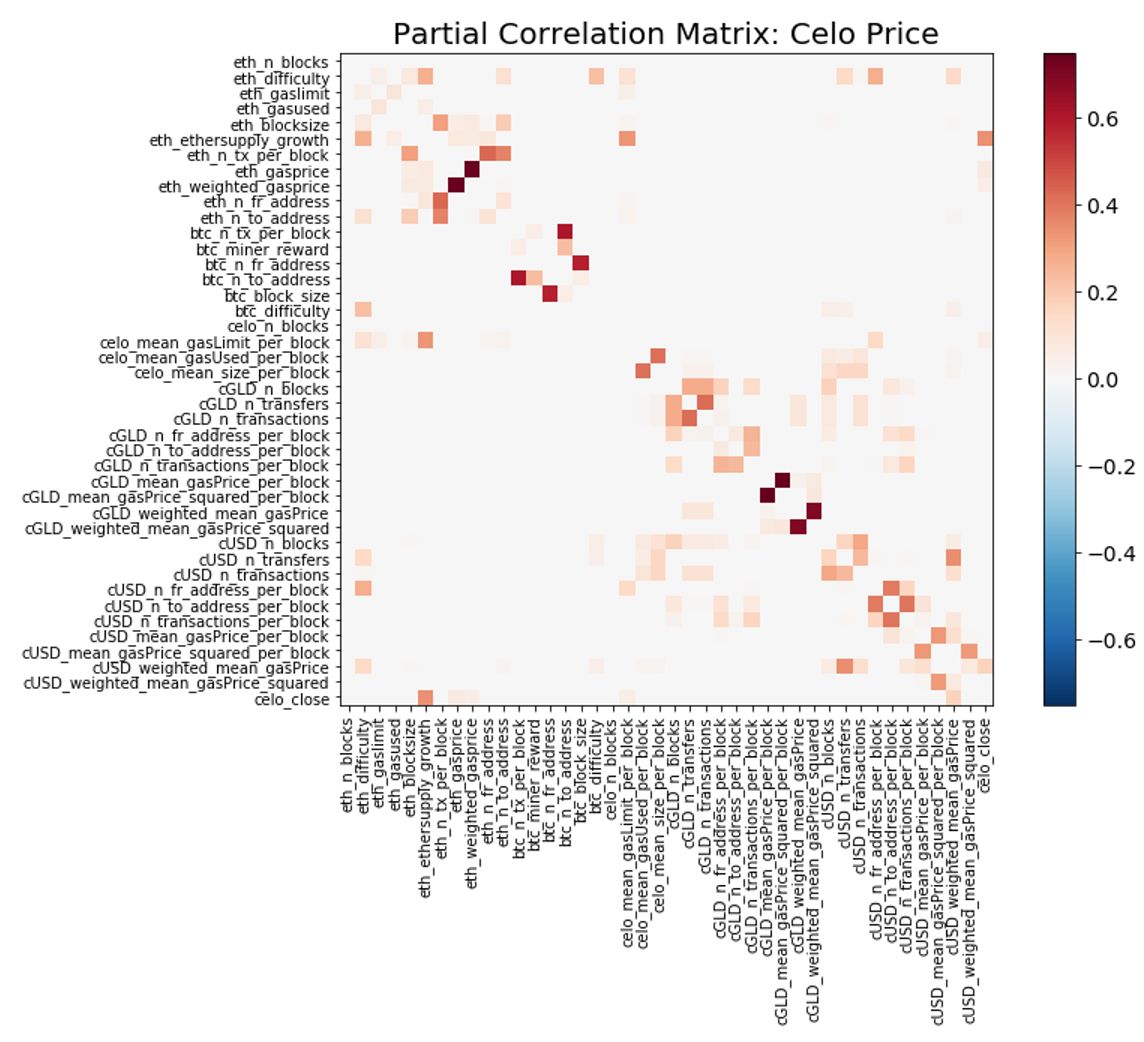}
	\caption{Partial correlation matrix from sparse inverse covariance estimation.}
	\label{fig:partial-corr-matrix-celo}
\end{figure}

\begin{figure}[H]
	\centering
	\includegraphics[width=\textwidth]{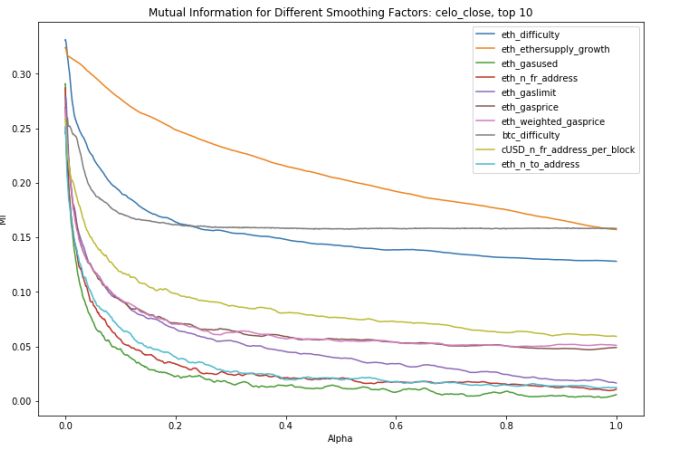}
	\caption{Mutual information of price data and features, with smoothing $\alpha$.}
	\label{fig:mutual-info-celo}
\end{figure}

The price recovery is generally poorer than for the ETH/USD price explored earlier. This is likely explained by the higher volatility of Celo compared to Ethereum as well as the smaller size of historical data available.

\begin{figure}[H]
	\centering
	\includegraphics[width=\textwidth]{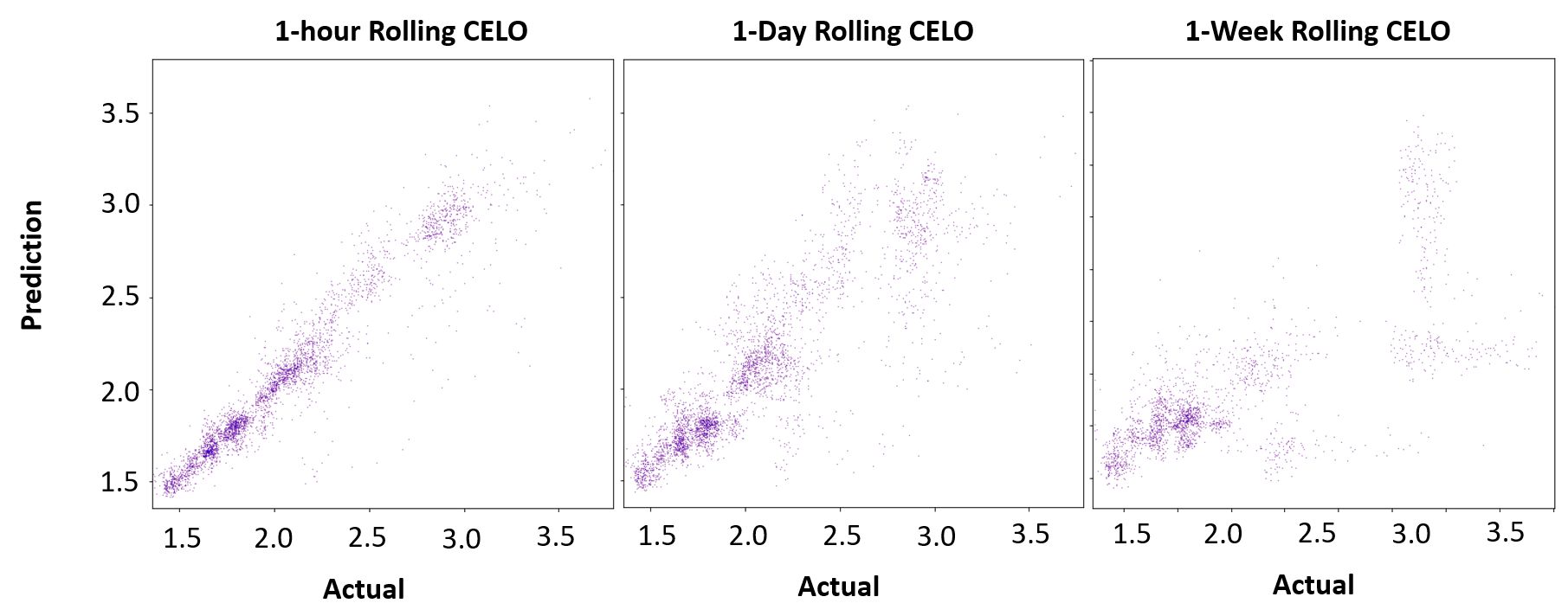}
	\caption{Recovered price vs actual for random forest with given retraining periods.}
	\label{fig:pred-retrainings-celo}
\end{figure}

\end{document}